\documentclass[apjl]{emulateapj}
\usepackage{apjfonts}
\newcommand*{\Msun}{\ensuremath{\mathrm{M_\odot}}}%
\newcommand*{\ML}{\ensuremath{M/L}}%
\newcommand*{\Vmax}{\ensuremath{V_{\mathrm{max}}}}%
\newcommand*{\Mpc}{\ensuremath{\mathrm{Mpc}}}%

\begin{document}

%% ------------------------------------------------------------------
%% TITLE
%% ------------------------------------------------------------------

\title{The stellar mass function of galaxies to \lowercase{$z \sim 5$} in
  the Fors Deep and GOODS-S fields}

\shorttitle{Stellar mass function at $0<z<5$}
\shortauthors{Drory et al.}

%% ------------------------------------------------------------------
%% AUTHORS
%% ------------------------------------------------------------------

\author{N.~Drory\altaffilmark{1},
  M.~Salvato\altaffilmark{3}, 
  A.~Gabasch\altaffilmark{3},
  R.~Bender\altaffilmark{2,3},
  U.~Hopp\altaffilmark{2},
  G.~Feulner\altaffilmark{2}}
\author{M.~Pannella\altaffilmark{3}}

\affil{$^1$ University of Texas at Austin, Austin, Texas 78712}
\email{drory@astro.as.utexas.edu}

\affil{$^2$ Universit\"ats--Sternwarte M\"unchen, Scheinerstra\ss
  e 1, D-81679 M\"unchen, Germany}
\email{hopp@usm.uni-muenchen.de}

\affil{$^3$ Max--Planck Institut f\"ur extraterrestrische Physik,
  Giessenbachstra\ss e, Garching, Germany}
\email{\{bender,gabasch,mara\}@mpe.mpg.de}

%% You can insert a short comment on the title page using the command below.
\slugcomment{Submitted to ApJ}

%% ------------------------------------------------------------------
%% ABSTRACT
%% ------------------------------------------------------------------

\begin{abstract}
  We present a measurement of the evolution of the stellar mass
  function (MF) of galaxies and the evolution of the total stellar
  mass density at $0 < z < 5$, extending previous measurements to
  higher redshift and fainter magnitudes (and lower masses). We use
  deep multicolor data in the Fors Deep Field (FDF; $I$-selected
  reaching $I_{AB} \sim 26.8$) and the GOODS-S/CDFS region
  ($K$-selected reaching $K_{AB} \sim 25.4$) to estimate stellar
  masses based on fits to composite stellar population models for 5557
  and 3367 sources, respectively.  The MF of objects from the
  $K$-selected GOODS-S sample is very similar to that of the
  $I$-selected FDF down to the completeness limit of the GOODS-S
  sample. Near-IR selected surveys hence detect the more massive
  objects of the same principal population as do $I$-selected surveys.
  We find that the most massive galaxies harbor the oldest stellar
  populations at all redshifts. At low $z$, our MF follows the local
  MF very well, extending the local MF down by a decade to
  $10^8$~\Msun.  Furthermore, the faint end slope is consistent with
  the local value of $\alpha \sim 1.1$ at least up to $z \sim 1.5$.
  Our MF also agrees very well with the MUNICS and K20 results at $z
  \lesssim 2$. The MF seems to evolve in a regular way at least up to
  $z \sim 2$ with the normalization decreasing by 50\% to $z=1$ and by
  70\% to $z=2$. Objects with $M > 10^{10}$~\Msun which are the likely
  progenitors of todays $L > L^*$ galaxies are found in much smaller
  numbers above $z \sim 2$.  However, we note that massive galaxies
  with $M > 10^{11}$~\Msun are present even to the largest redshift we
  probe.  Beyond $z \sim 2$ the evolution of the mass function becomes
  more rapid.  We find that the total stellar mass density at $z=1$ is
  50\% of the local value.  At $z=2$, 25\% of the local mass density
  is assembled, and at $z=3$ and $z=5$ we find that at least 15\% and
  5\% of the mass in stars is in place, respectively. The number
  density of galaxies with $M > 10^{11}\,\Msun$ evolves very similarly
  to the evolution at lower masses. It decreases by 0.4~dex to $z \sim
  1$, by 0.6~dex to $z \sim 2$, and by 1~dex to $z \sim 4$.
\end{abstract}

%% KEYWORDS
%%
\keywords{surveys --- cosmology: observations --- galaxies: mass
  function --- galaxies: evolution --- galaxies: fundamental parameters}

%% ------------------------------------------------------------------
%% INTRODUCTION
%% ------------------------------------------------------------------

\section{Introduction}\label{sec:introduction}

The stellar mass of galaxies at the present epoch and the build-up of
stellar mass over cosmic time has become the focus of intense research
in the past few years.

Generally, this kind of work relies on fits of multi-color photometry
to a grid of composite stellar population (CSP) models to determine a
stellar mass-to-light ratio, since large and complete spectroscopic
samples of galaxies (at $z>0$) are not yet available.

In the local universe, results on the stellar mass function (MF) of
galaxies were published using the new generation of wide-angle surveys
in the optical (Sloan Digital Sky Survey; SDSS, \citealp{SDSS}; 2dF,
\citealp{2dF}) and near-infrared (Two Micron All Sky Survey; 2MASS,
\citealp{TwoMASS}). \citet{2dF01} combined data from 2MASS and 2dF to
derive the local stellar MF, \citet{BMKW03} used the SDSS and 2MASS to
the same end.

At $z > 0$, a number of authors studied the stellar mass density as a
function of redshift
\citep{BE00,MUNICS3,Cohen02,DPFB03,Fontanaetal03,Rudnicketal03}
reaching $z \sim 3$. It appears that by $z \sim 3$, about 30\% of the
local stellar mass density has been assembled in galaxies, and at $z
\sim 1$, roughly half of the local stellar mass density is in place.
This seems to be in broad agreement with measurements of the star
formation rate density over the same redshift range.

Others investigated the evolution of the MF of galaxies
\citep{MUNICS6,K20-04} to $z \sim 1.5$, finding a similar decline in
the normalization of the MF. However, it is possible that galaxies
evolve differently in number density depending on their morphology.

In this letter, we extend the measurement of the stellar MF and its
integral, the total stellar mass density, to $z \sim 5$. We describe
the data set in Sect.~\ref{sec:galaxy-sample} and present the method
used to derive stellar masses in Sect.~\ref{sec:deriving-masses}. We
discuss the results on the stellar MF in
Sect.~\ref{sec:mass-function}. Finally, we discuss the total stellar
mass density and the number density of massive galaxies in
Sect.~\ref{sec:mass-density}.

Throughout this work we assume $\Omega_M = 0.3$, $\Omega_{\Lambda} =
0.7, H_0 = 70\ \mathrm{km\ s^{-1}\ Mpc^{-1}}$. Magnitudes are given in
the AB system.

%% ------------------------------------------------------------------
%% THE GALAXY SAMPLE
%% ------------------------------------------------------------------

\section{The galaxy sample}\label{sec:galaxy-sample}

This work is based on data from two deep field surveys having
multicolor photometry and rich followup spectroscopy, the Fors Deep
Field (FDF; \citealp{FDF1,FDF2}) covering the UBgRIzJKs bands in 40
arcmin$^2$ and the GOODS-S/CDFS field \citep{GOODS1} covering the
UBVRIJHKs bands in 50 arcmin$^2$. This sample is identical to the one
used in \citet{Gabasch04a} to study the global star formation rate to
$z \sim 5$.

The FDF photometric catalog is published in \citet{FDF1} and we use
the I-band selected subsample covering the deepest central region of
the field as described in \citet{Gabasch04}.  This catalog lists 5557
galaxies down to $I \sim 26.8$. The latter also discusses the I-band
selection and shows that this catalog misses at most 10\% of the
objects found in ultra-deep K-band observations found by
\citet{FIRES03}.

Photometric redshifts for the FDF are calibrated against 362
spectroscopic redshifts up to $z \sim 5$ and have a accuracy of
\mbox{$\Delta z / (z_{spec}+1) \le 0.03$} with only $\sim 1$\%
outliers. The method and this calibration are presented and discussed
in \citet{Gabasch04}.

Our K-band selected catalog for the GOODS-S/CDFS field is based on
the publicly available 8 $2.5\times2.5$ arcmin$^2$ J, H, Ks ISAAC
images, taken at ESO/VLT with seeing in the range $0.4'' - 0.5''$.
The U and I images are from ESO GOODS/EIS public survey, while the B V
R images are taken from the Garching-Bonn Deep Survey. These datasets
are extensively discussed in \citet{Arnouts01} and \citet{Schirmer03},
respectively.

The data for the GOODS field were analyzed in a very similar way to
the data of the FDF.  The objects were detected in the K-band images
closely following the procedure used for the FDF I-band detection. A
detailed description of the procedure as well as the catalogs can be
found in Salvato et al.\ (2004, in preparation).  We detect 3367
sources in the K-band for which we measure fixed aperture and
Kron-like total magnitudes in all bands.  Of these, 2 objects are
detected only in K, and 36 objects are detected only in the Ks, H, and
J band images. Number counts match the literature values down to
$K\approx 25.4$, which is the completeness limit of the catalog. We
computed photometric redshifts using the same method and SED template
spectra as for the FDF.  A comparison with the spectroscopic redshifts
of the VIMOS team and the FORS2 spectra shows that the photometric
redshifts have an accuracy $\Delta z /(z_{spec}+1) \le 0.056$.

%% ------------------------------------------------------------------
%% DERIVING STELLAR MASSES
%% ------------------------------------------------------------------

\section{Deriving stellar masses}\label{sec:deriving-masses}

We derive stellar masses (the mass locked up in stars) by comparing
multi-color photometry to a grid of stellar population synthesis
models covering a wide range in parameters, especially star formation
histories (SFHs).  This method is described in detail and compared
against spectroscopic and dynamical mass estimates in Drory, Bender,
\& Hopp \citeyearpar{DBH04}, hence we only briefly outline the method
here.

We base our model grid on the \citet{BC03} models.  We parameterize
the possible SFHs by a two-component model, a main component with a
smooth analytical SFH modulated by a burst of star formation.  The
main component is parameterized by a SFH of the form $\psi(t) \propto
\exp(-t/\tau)$, with $\tau \in [0.1, \infty]$~Gyr and a metallicity of
$-0.6 < \mathrm{[Fe/H]} < 0.3$.  The age, $t$, (defined as the time
since star formation began) is allowed to vary between 0.5~Gyr and the
age of the universe at the object's redshift. This component is
linearly combined with a burst modeled as a 100~Myr old constant star
formation rate episode of solar metallicity. We restrict the burst
fraction, $\beta$, to the range $0 < \beta < 0.15$ in mass (higher
values of $\beta$ are degenerate and unnecessary since this case is
covered by models with a young main component).  We adopt a Salpeter
IMF truncated at 0.1 and 100~\Msun\ for both components. A different
slope will change only the mass scale.  However, if the IMF depends on
the mode of star formation the shape of the MF can be affected.

\begin{figure}[t]
  \includegraphics[width=8cm]{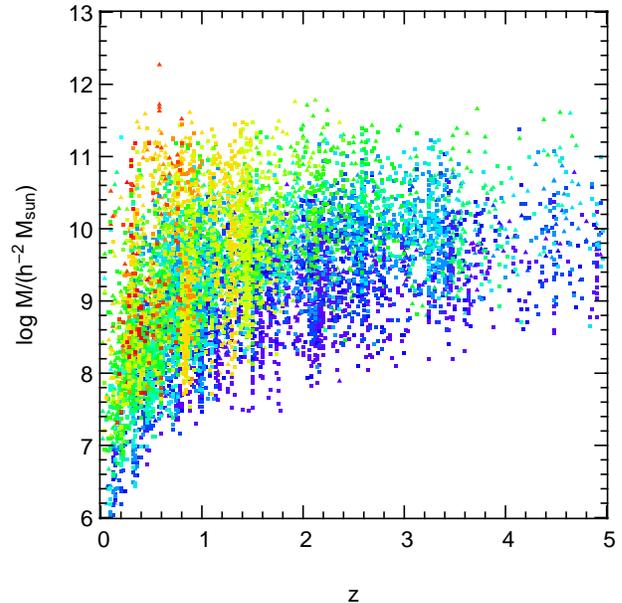}
  \caption{Stellar masses vs.\ redshift in the FDF (squares) and in 
    GOODS-S (triangles). The luminosity-weighted mean ages of the stellar
    populations as determined by the CSP fit are coded by colors ranging
    from red for old populations (age $\gtrsim 5$~Gyr) to blue for young
    populations (age $\lesssim 1.5$~Gyr). \label{f:Mz}}
\end{figure}

Additionally, both components are allowed to exhibit a different and
variable amount of extinction by dust.  This takes into account the
fact that young stars are found in dusty environments and that the
starlight from the galaxy as a whole may be reddened by a (geometry
dependent) different amount. In fact, \citet{SMSS04} find that the
Balmer decrement in the SDSS sample is independent of inclination,
which, on the other hand, is driving global extinction (\citealp[see,
e.g.,][]{TPHSVW98}).

We compute the full likelihood distribution on a grid in this
6-dimensional parameter space ($\tau, \mathrm{[Fe/H]}, t, A_V^1,
\beta, A_V^2$), the likelihood of each model being $\propto
\exp(-\chi^2/2)$. To compute the likelihood distribution of \ML, we
weight the \ML\ of each model by its likelihood and marginalize over
all parameters. The uncertainty in \ML\ is obtained from the width of
this distribution, which is on average between $\pm 0.1$ and $\pm
0.3$~dex at 68\% confidence level. The uncertainty in mass has a weak
dependence on mass (increasing with lower $S/N$ photometry) and much
of the variation is in spectral type: early-type galaxies have more
tightly constrained masses than late types. At high redshift, however,
the uncertainty in \ML\ grows as objects drop-out in the blue bands
and stellar populations become younger: the typical uncertainty at $z
\sim 3$ is 0.6~dex and becomes as large as 1~dex at the highest
redshifts we probe.

In Fig.\ref{f:Mz} we show the distribution of galaxies in the mass
vs.\ redshift plane for the FDF (squares) and GOODS-S (triangles). In
addition, we code the age of each galaxy (using the best-fitting
model) in colors ranging from blue for young (age $\lesssim 1.5$~Gyr)
to red for old stellar populations (age $\gtrsim 5$~Gyr).

The distribution of objects from the $K$-selected GOODS-S sample is
very similar to the distribution of the $I$-selected FDF down to the
completeness limit of the GOODS-S sample, which is $\sim 1$~dex
shallower in limiting mass. This indicates that present optical and
near-IR surveys are unlikely to have missed a substantial population
of massive objects, with the possible exception of heavily
dust-enshrouded sources which may escape detection in both optical and
near-IR surveys.

\begin{figure*}
  \includegraphics[width=\textwidth]{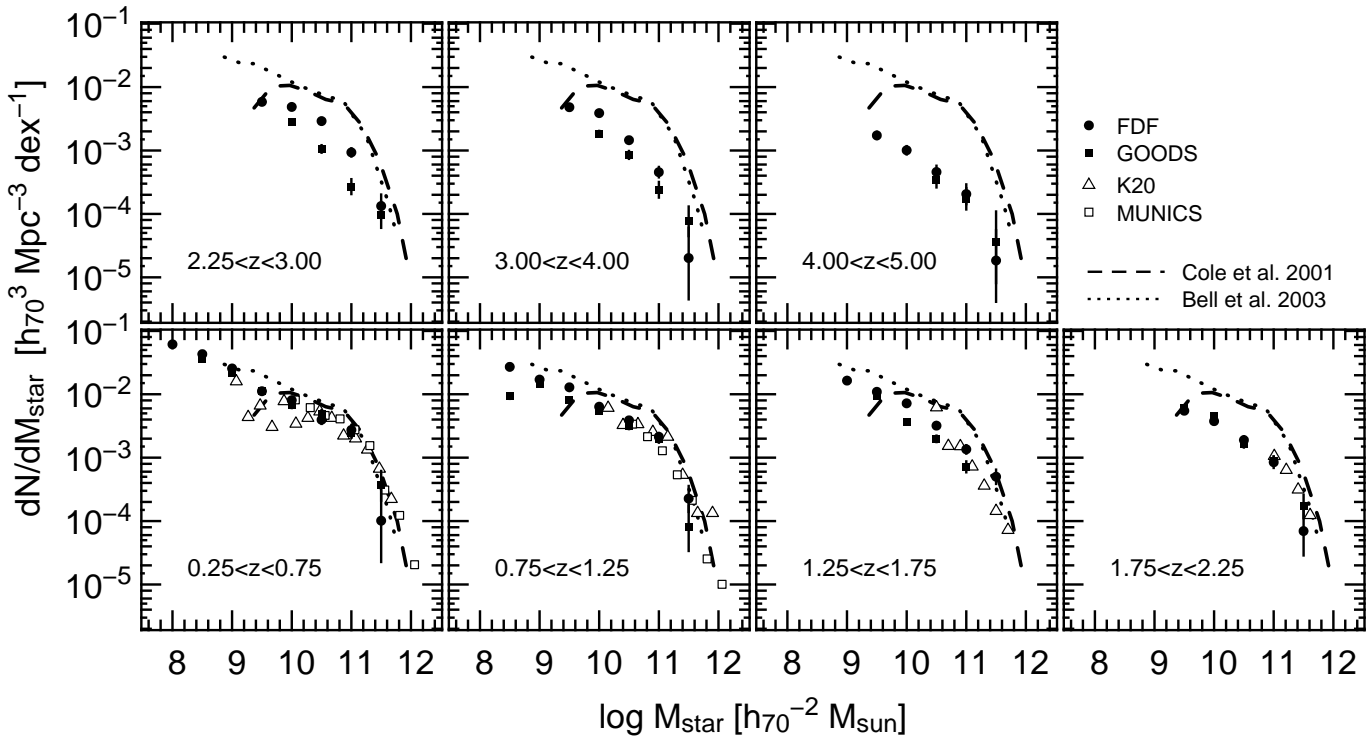}
  \caption{The stellar mass function (MF) as a function of redshift. 
    The local MF of \citet{BMKW03} and \citet{2dF01} are shown in all
    redshift bins for comparison. Results from the MUNICS survey
    \citep{MUNICS6} and the K20 survey \citep{K20-04} at $z \lesssim
    2$ are also shown. \label{f:mf}}
\end{figure*}

A striking feature of Fig.\ref{f:Mz} is that the most massive galaxies
harbor the oldest stellar populations at all redshifts. There always
exist galaxies which are old (relative to the age of the universe at
each redshift) but less massive, yet the most massive objects at each
redshift are never among the youngest.

%% ------------------------------------------------------------------
%% THE STELLAR MASS FUNCTION
%% ------------------------------------------------------------------

\section{The stellar mass function}\label{sec:mass-function}

Fig.~\ref{f:mf} shows the \Vmax-weighted mass function in seven
redshift bins from $z=0.25$ to $z=5$. For comparison, we also show the
local mass function \citep{BMKW03,2dF01} and the MFs to $z \sim 1.2$
of MUNICS \citep{MUNICS6} and to $z \sim 2$ by the K20 survey
\citep{K20-04}.

The MF of objects from the $K$-selected GOODS-S sample is
very similar to that of the $I$-selected FDF down to the
completeness limit of the GOODS-S sample (with the exception of a
possible slightly lower normalization at $z \gtrsim 2.5$ by about 10\%
to 20\%).  This shows that near-IR selected surveys at high redshift
essentially detect the more massive objects of the same principal
population as do optically ($I$-band) selected surveys (see also
Sect.~3 and Fig.~1 in \citealp{Gabasch04}).  It remains to be seen
what fraction of massive galaxies (and total stellar mass density)
might be hidden in dusty starbursts which appear as sub-mm galaxies.

In our lowest redshift bin, $z \sim 0.5$, the MF follows the local MF
very well. The depth of the FDF ($I \sim 26.8$) allows us to extend
the faint end of the MF down to $10^8$~\Msun, a decade lower in mass
than in \citet{BMKW03}, with no change of slope. Furthermore, the
faint end slope is consistent with the local value of $\alpha \sim
1.1$ at least to $z \sim 1.5$. Our mass function also agrees very well
with the MUNICS and K20 results at $z \lesssim 2$.

The MF seems to evolve in a regular way at least up to $z \sim 2$ with
the normalization decreasing by 50\% to $z=1$ and by 70\% to $z=2$,
with the largest change occurring at masses of $M ~ 10^{10}$~\Msun.
These likely progenitors of todays $L > L^*$ galaxies are found in
much smaller numbers above $z \sim 2$. However, we note that massive
galaxies with $M > 10^{11}$~\Msun are present even to the largest
redshift we probe (albeit in smaller numbers). Beyond $z \sim 2$ the
evolution becomes more rapid.

It is hard to say whether the difference between the FDF and the
GOODS-S field visible at $z \sim 3$ in Fig.~\ref{f:mf} is significant.
In fact, \citet{Gabasch04a} find not much difference in the rest-frame
UV luminosity function in the very same dataset. We think it might be
related to the shallower near-IR data in the FDF compared to GOODS-S.
Less reliable information on the rest-frame optical colors at young
mean ages might in fact lead to an overestimation of the stellar mass.
This would also explain why the two MFs become similar again at even
higher $z$ when the near-IR data in GOODS-S reach their limit, too.
This would mean that the mass density in the FDF at $z \gtrsim 2.5$
might be overestimated (see below).

%% ------------------------------------------------------------------
%% THE TOTAL STELLAR MASS DENSITY
%% ------------------------------------------------------------------

\section{Stellar mass density and number densities}
\label{sec:mass-density}

\begin{figure}
  \includegraphics[width=8cm]{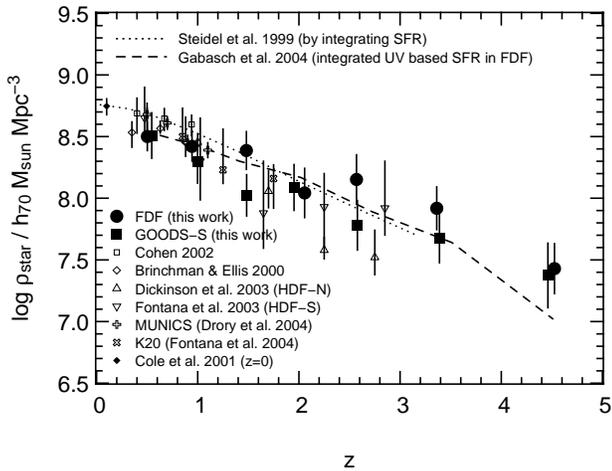}
  \caption{The total stellar mass density of the universe as a function 
    of redshift. We also show the integral of the star formation rate
    as a dotted line and dashed line.
    \label{f:md}}
\end{figure}

Fig.~\ref{f:md} shows the evolution of the total stellar mass density
along with data from the literature, extending the available data to
$z \sim 5$
\citep{2dF01,BE00,MUNICS3,Cohen02,DPFB03,Fontanaetal03,K20-04,MUNICS6}.

We compute the total stellar mass density by directly summing up
contributions from all objects in both fields (we obtain very similar
values by means of fitting Schechter functions to the data in
Fig.~\ref{f:mf}).  We find that the stellar mass density at $z=1$ is
50\% of the local value as determined by \citet{2dF01}.  At $z=2$,
25\% of the local mass density is assembled, and at $z=3$ and $z=5$ we
find that at least 15\% and 5\% of the mass in stars is in place,
respectively. Fig.~\ref{f:md} shows the results from both fields
separately, however.

We also show the integral of the star formation rate determined by
\citet{SAGDP99} as a dotted line. Furthermore, the dashed line shows
the same quantity determined from the rest-frame UV luminosity
function of in same dataset used here \citep{Gabasch04a}. We find
these measurements in agreement with each other, and with the mass
densities derived here and in the literature before. However, the mass
densities do show considerable scatter especially at redshifts above
$z \sim 1.5$.  However, as discussed above, the mass density in the
FDF might be overestimated, hence reducing the scatter between our two
fields.  The FDF also lies above the average literature values in the
redshift range in question, while the GOODS-S data seem to be in
better agreement with previous measurements.

\begin{figure}
  \includegraphics[width=8cm]{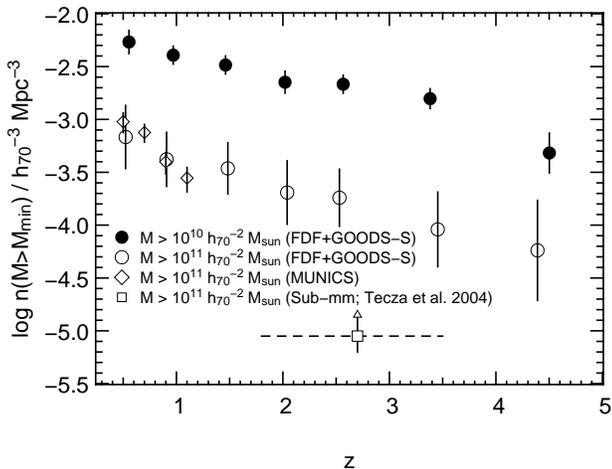}
  \caption{The number density of galaxies with stellar masses 
    $M > 10^{10} \Msun$ and $M > 10^{11} \Msun$. Results from the
    MUNICS survey as well as results from Sub-mm studies are also
    shown.
    \label{f:nd}}
\end{figure}

The fact that the stellar mass densities and the integrated star
formation rates give consistent results is encouraging. However,
dust-enshrouded starbursts at high redshift would be missing from both
the SFR and the MF. However, if the number density of massive sub-mm
galaxies derived by \citet{Teczaetal04} of $8.9_{-3.3}^{+5.3}\times
10^{-6}\,\Mpc^{-3}$ at $1.8 < z < 3.6$ is typical of these systems,
they do not contribute significantly to the mass density at these
redshifts (see below and Fig.\ref{f:nd}).

Finally, Fig.~\ref{f:nd} addresses the number density of massive
systems as a function of redshift. We show the numbers of systems with
$M > 10^{10}\,\Msun$ and $M > 10^{11}\,\Msun$ as full and open
symbols, respectively. The number density of massive sub-mm galaxies
estimated by \citet{Teczaetal04} is also shown. The most striking
features of Fig.~\ref{f:nd} is that the number density of the most
massive systems shows evolution which is very similar to the evolution
of the number density at lower masses over this very large redhsift
range. Massive systems are present at all redshifts we probe, their
number density decreasing by $0.4\pm 0.2$~dex to $z \sim 1$, by
$0.6\pm 0.3$~dex to $z \sim 2$, and by $1 \pm 0.45$~dex to $z \sim 4$.

%% ------------------------------------------------------------------
%% ACKNOWLEDGMENTS
%% ------------------------------------------------------------------

\acknowledgments

This work was partly supported by the Deutsche Forschungsgemeinschaft,
grant SFB 375 ``Astroteilchenphysik.'' N.D.\ acknowledges support by
the Alexander von Humboldt Foundation. This work makes use of data
obtained by the ESO GOODS/EIS project using the Very Large Telescope
at the ESO Paranal Observatory under Program ID(s): LP168.A-0485.

%% ------------------------------------------------------------------
%% REFERENCES
%% ------------------------------------------------------------------

%\bibliography{literature} \bibliographystyle{apj}

\end{document}